# Auxiliary Optics for meV-Resolved Inelastic X-Ray Scattering at SPring-8: Microfocus, Analyzer Masks, Soller Slit, Soller Screen, and Beam Position Monitor

Alfred Q.R. Baron[1,a], Daisuke Ishikawa[1,2], Hiroshi Fukui[1,3] and Yoichi Nakajima[1,4]

[1]*Materials Dynamics Laboratory, RIKEN SPring-8 Center, 1-1-1 Kouto, Sayo, Hyogo 679-5148, JAPAN*
[2]*SPring-8/JASRI, 1-1-1 Kouto, Sayo, Hyogo 6789-5198, JAPAN*
[3]*Graduate School of Material Science, University of Hyogo, 3-2-1 Kouto, Kamigori, Hyogo 678-1297 JAPAN*
[4]*Department of Physics, Kumamoto University, Kumamoto 860-8555, JAPAN*

[a]Corresponding author: baron@spring8.or.jp

**Abstract.** This paper discusses several optical elements now in use at BL43LXU, the RIKEN Quantum NanoDynamics Beamline, of the RIKEN SPring-8 Center. BL43LXU is dedicated to meV-resolved inelastic x-ray scattering (IXS) using spherical analyzers operating between 15.8 and 25.7 keV. The work described here is relevant for setups on the high-resolution spectrometer (10m two-theta arm) with resolution between 2.8 and 0.8 meV. Specific optics discussed include a multilayer Kirkpatrick-Baez (KB) mirror pair that focuses the full (~1x3mm$^2$) beam at 17.79 keV to a 4.4 x 4.1 μm$^2$ spot with ~60% throughput, two different types of Soller slits that help reduce backgrounds, masks for the analyzers that allow increased solid angle to be collected while preserving momentum resolution, and a diamond quadrant beam position monitor (BPM). These elements have been used for experiments in extreme conditions with diamond anvil cells, and liquid measurements at low momentum transfers, among other work.

## INTRODUCTION

Inelastic x-ray scattering (IXS) with meV resolution offers the opportunity to probe atomic dynamics (the dynamic structure factor, $S(\mathbf{Q},\omega)$) in cases when it is not easily accessible using other methods. Specific examples include the measurement of small (~10 micron) samples and the measurement of large energy transfers at small momentum transfers. Thus meV-IXS is often used for investigations of samples in extreme (high pressure, high temperature) conditions in diamond anvil cells (DACs) and investigation of liquids, where the low-Q region can have unique information. The present paper describes some auxiliary instrumentation used to facilitate these measurements, as implemented at the RIKEN Quantum NanoDynamics beamline, BL43LXU [1], of the RIKEN SPring-8 Center. A broader introduction to IXS experiments and optics may be found in [2].

The motivation for this paper is primarily to show what has been done and found to be useful. It is not intended to be an in-depth look at the background for the more typical components (Kirkpatrick-Baez (KB) mirrors and Soller slits) so much as to provide one set of operating parameters. In particular, KB mirrors are in use in many beamlines, but we were not able to find details on any that were very close in parameter space to what we considered. Meanwhile, we know of no cases where tailored analyzer masks, as discussed here, or Soller slits, or what we call a Soller screen, have been used for IXS. The beam position monitor is mostly a combination of commercial components (with some specialization of the mechanics) and we show reasonable, even easy, usage at the 0.01 mm level, as is sufficient for the way it is used here.

# ELLIPTICAL MULTILAYER KB MIRROR PAIR AT 17.793 KEV

Focusing the x-ray beam is highly desirable to increase count rates when small samples are measured. At first glance, having a smaller beam is generally better, as most experiments using a large beam can also be done with a small beam, but not the reverse. However, small beam sizes also usually have increased divergence (i.e.: conserved brilliance), and strong focusing optics often lead to limited clear aperture (free space, without optics) around the sample, reduced throughput, and higher radiation damage, all of which may negatively impact experiments. Thus the focal spot size for an experiment is a compromise between competing issues. For the present case we targeted a 5 μm beam size (unless otherwise stated, all sizes mentioned will be full width at half maximum, FWHM) as being a reasonable compromise between what was needed for high-pressure studies, up to ~300 GPa, and the other issues mentioned. Previously the SPring-8 IXS beamlines have used compound focusing as this was easiest to implement: the addition of an optic after the usual bent cylindrical mirror was used to reduce the beam from ~50 um to ~15 μm [3]. However, it is difficult to go to a smaller size in that compound geometry as (a) the compound focusing tends to limit the demagnification possible for a given sample clear aperture, and (b) the focusing of the second component must overcome the errors of the first optic. These issues led us to implement a single focusing elliptical Kirkpatrick-Baez (KB) mirror pair ("single focusing" means just one focusing element in each direction transverse to the beam). However, this means that there is no upstream component to help reduce the beam size so the installed optic must accept the full beam size (~1x3mm$^2$, full width) near the sample, ~100m from the source to avoid large losses. This forces operation at larger (~1 degree) grazing angles to allow reasonable size optics, which in turn basically requires a multilayer optic. As ellipsoidal optics have not yet progressed to a level sufficient for this problem, we use an elliptical KB mirror pair with multilayer coatings.

The parameters for the mirror pair are given in table 1. Notably the vertical focusing is stronger than the horizontal as is unusual at a SR facility. However the beamline has 7 vertically scattering reflections upstream of the KB pair (2 mirrors and 5 crystal reflections), so it was known that the effective source size in the vertical was degraded, so strong focusing was chosen to compensate for this. The mirror substrates were polished by J-Tec, and they successfully achieved all critical specifications. The multilayer was deposited by Rigaku Innovative Technologies (RIT, previously Osmic) and consisted of 200 layers of $B_4C/Mo$ (approximately 2:1 thickness ratio) with a 95 Å SiC cap layer. The multilayer on the vertical focusing mirror was graded from d=2.1 to 2.8 nm and the horizontal from 2.2 to 2.7 nm. The slope error specification was set to limit the nominal contribution from this to less than 1μm FWHM. The relatively good, 0.2 nm rms, roughness specification was to keep good reflectivity from the multilayer.

Figure 1 shows the results from measuring the KB pair. The reflectivity was found to be good over-all, but in each case, there is a slight worsening of the reflectivity as the d-spacing becomes smaller (translation <0), consistent with calculation. But, in both cases the average reflectivity over the beam spot was > 75% and the throughput of the pair was ~60% at 17.793 keV. The measured beam spot size can be seen in the lower panel of fig. 1, and the size, 4.4 μm V x 4.1 μm H, is both within the desired specification and reasonable given the concerns mentions above, namely, the perturbation of the vertical beam response by other optical elements, and the presence of multiple source points. We note, for completeness and reference, that the beam sizes given by differentiation using the PyMca code are about 10% larger than those from our fits, with a smoother looking derivative plot.

|  | Vertical | Horizontal |
|---|---|---|
| **Operating Energy** | 17793 (+-3) eV | |
| **Distance to Sample (q)** | 0.2 m | 0.4 m |
| **Distance to Source (p)** | 109.8 m | 109.6 m |
| Desired Spot Size (FWHM) | <5 μm FWHM | <5 μm FWHM |
| Nominal Demagnification | ~500 | ~270 |
| Nominal Source Size | < 0.1 mm | ~0.6 mm |
| Ideal Focus | < 1 μm | ~2.2 μm |
| Est. Multi-Source Contribution | < 1 μm | ~1.5 μm |
| Est. Slope Error Contribution ~2 x 2.35 x rms x q | <1 μm | <1 μm |
|  |  |  |
| **Substrate Material** | Single Crystal Silicon | |
| **Substrate Outer Dimensions (L x W x H)** | 80 x 40 x 15 mm$^3$ | 250 x 40 x 30 mm$^3$ |
| **Grazing Angle of Incidence at Mirror Center** | 14.00 (+-0.02) mrad | |
| **Mirror Active Length** | >70 mm | >240 mm |
| **Mirror Active Width** | > 10 mm | > 6 mm |
| Acceptance | > 0.95 mm | > 3.35 mm |
| **Roughness** | < 0.2 nm rms | |
| **Longitudinal Slope Error (rms)** | < 1 μrad | <0.5 μrad |
| **Transverse Slope Error (rms)** | < 20 μrad | <10 μrad |
| **Reflectivity at operating energy** | > 70 % | |

**TABLE 1.** Design parameters for the KB mirror pair (partial list). Boldface quantities are primary while others are secondary and are included for convenience/reference.

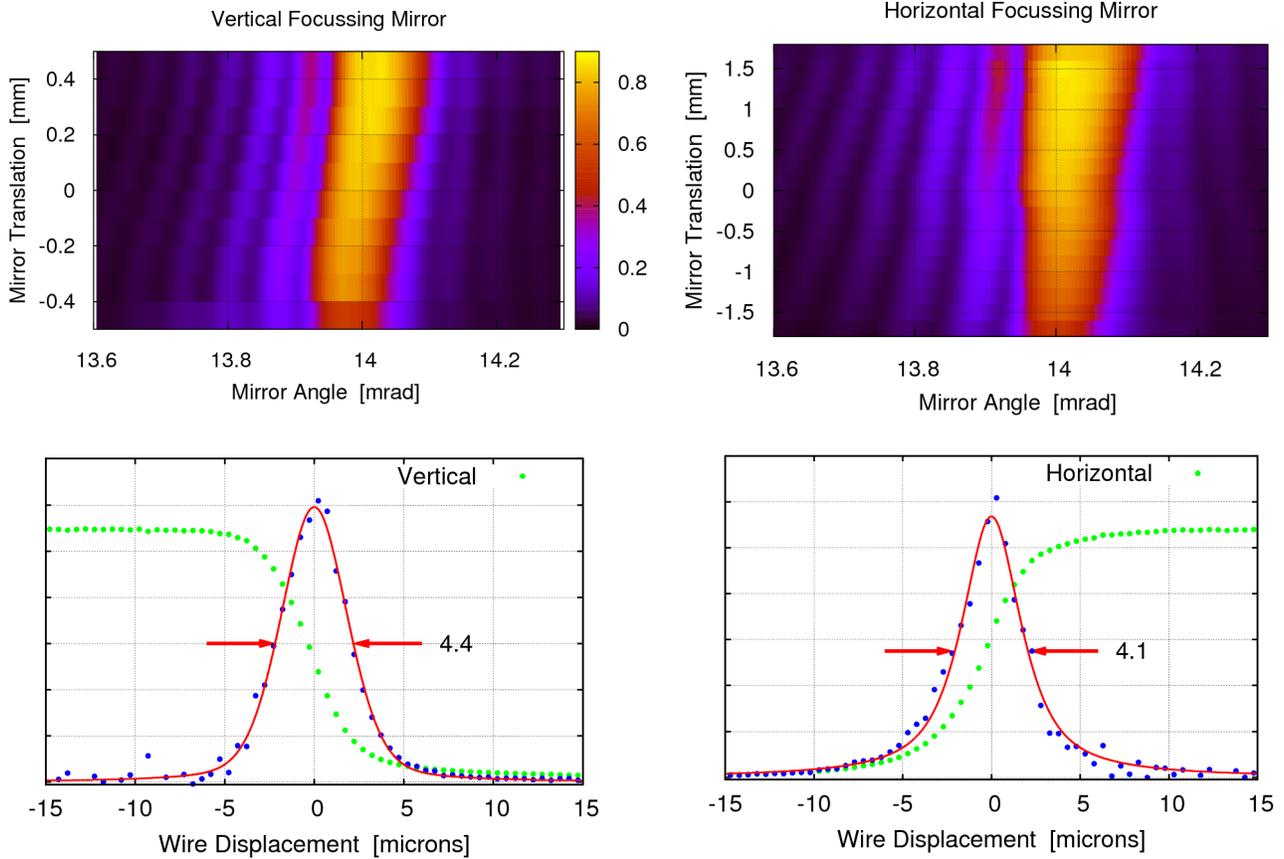

**FIGURE 1. KB Mirror Performance.** The top panels give the measured reflectivity of each mirror as a function of position and grazing angle when the mirror is scanned through a small beam. The main Bragg peak at 14 mrad and the usual subsidiary maxima are easily visible. The lower panels show the beam profiles measured using a wire scan with the calculated derivative and a pseudo-Voigt fit. The numbers give the FWHM from the fit.

## ANALZYER MASKS FOR SMALL MOMENTUM TRANSFERS

The BL43LXU high-resolution spectrometer presently operates, mostly, using a 6x4 array of spherical analyzers located 9.8 m from the sample on a large two-theta arm, as can be seen in fig. 2. The acceptance of these analyzers is controlled by a set of motorized "venetian blind" [4] slits located about 9m from the sample that can be set to any rectangular size between 3x3 mm$^2$ and 80x85 mm$^2$, with, always, an identical size set for all analyzers (e.g. one motor for a common vertical gap control and one for a common horizontal gap). This geometry is effective for measuring crystalline samples (see discussion in [2]). However, for disordered materials at low momentum transfers one often desires an annular acceptance that corresponds to a constant momentum transfer: a rectangular slit is not optimized for fixed momentum transfer, |Q|, resolution at small Q. We therefore implemented a series of analyzer masks

specifically for low-Q measurements of disordered materials, with the goal of improving rates without sacrificing momentum resolution.

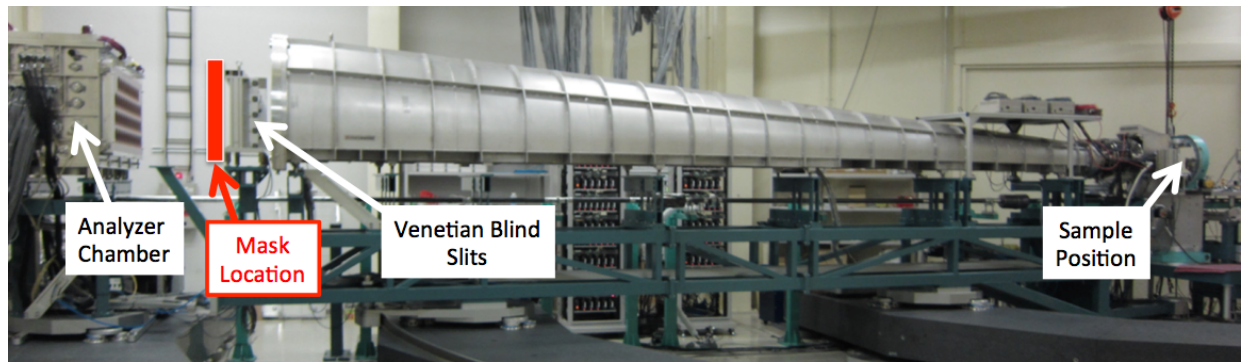

**FIGURE 2. Photo of the 10m Two-Theta arm at BL43LXU.** The sample position is the center of the Huber 512 Eulerian Cradle at the far right and the analyzers are kept inside the vacuum chamber at the far left. The air gap between the long flight path and the analyzer chamber (both built by Ayumi Industries) allows easy installation of the masks on the venetian blind slit (Kohzu) at 9m from the sample.

The masks were wire-cut in 3 mm aluminum (see fig. 3), and then covered by Pb tape and then Kapton tape, as seen in the photograph in fig. 3. Two sets were fabricated, one, mask set A, is designed to have the incident beam between the first two analyzers and collect momentum transfers down to ~0.5 nm$^{-1}$, simultaneously taking data at two small momentum transfers. The other, set B, operates with the first analyzer at slightly higher momentum transfers, >1 nm$^{-1}$, with the beam just to one side of the entire array. Both mask sets are designed to work with the Soller slit described in the next section. The shapes were chosen so the momentum resolution (3.5 times the rms deviation from the mean) at most momentum transfers was < 10% of the momentum transfer, with a worst case ~20% acceptance at the smallest |Q|. Calculations, such as those shown in fig. 3, were done to confirm the shape and magnitude of the each mask's |Q| acceptance. Eventually we expect to improve IXS modeling to explicitly include the calculated momentum resolution using continuous parameters (e.g. the slope of some mode dispersion, not just the mode frequency at one Q) and fit spectra at several momentum transfers simultaneously to a common parameter set.

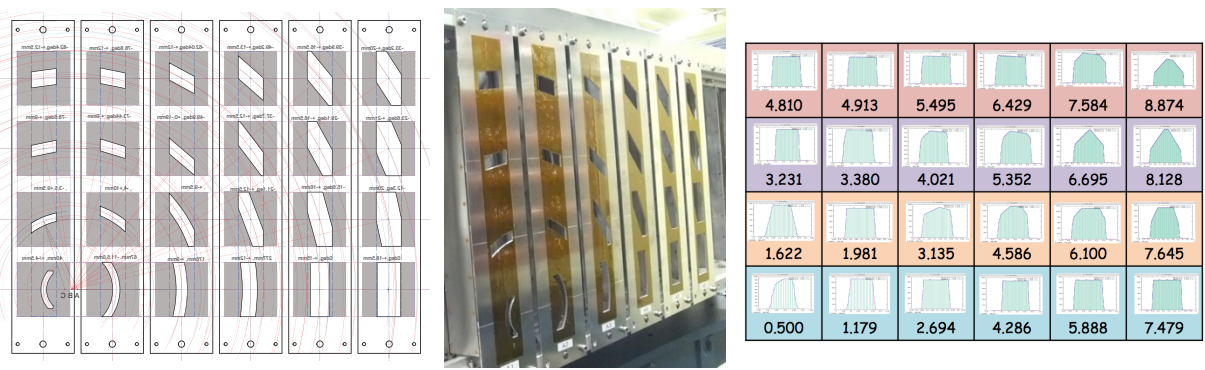

**FIGURE 3. Analyzer mask set "A".** The left panel shows the design drawing, the middle a photo of the masks installed on the pins behind the venetian blind slits, and the right shows the calculated momentum acceptance profile of all the masks - intensity vs momentum transfer - in one operational configuration with the numbers giving the central (centroid of) the momentum transfer of each analyzer in nm$^{-1}$. These are mostly reasonably flat-topped. Note that the last (right-hand) panel is plotted as seen from the sample (while the first two are as seen from the analyzers) so the plot for the lowest momentum transfer is in the lower right. The scale of the photo can be understood by noting the spacing between the mask centers is about 110 mm in each direction.

# SOLLER SLIT & SCREEN

Soller slits are a well-known and early [5], but still useful, method of reducing backgrounds in scattering experiments by limiting the detector (analyzer) field of view to a narrow region around the sample. Especially for IXS with spherical analyzers, there is a discrete separation of beams to different analyzers (see above) so it is natural to consider Soller slits matching the periodicity of the analyzers. However, Soller slits can be tricky to implement as, to be most effective, one would like small spacing between the different channels of the Soller slit. In the present case, the center-to-center spacing of the analyzer channels is 120 mm at 9.8m, corresponding to 12.2 mrad or 0.702 degrees. Here we discuss two geometries, one with a "conventional" Soller slit with an entrance window 68 mm from the sample (so a center-to-center channel spacing of 0.830 mm) and then what we call a Soller "screen" that starts 5 mm from the sample so the analyzer center-to-center spacing is 0.061 mm. In the first case, the conventional Soller slit is designed to operate with the masks mentioned above. Separately, though not yet tested, this Soller slit is also expected to be useful for looking at crystals at larger momentum transfers, to e.g., reduce backgrounds from the beryllium cap used in many refrigerators. In the second, 5mm, case, due to limits in fabrication technology, a pair of screens (without foils between channels), is used to collect *every other*[*] analyzer channel, with, effectively, half the channels sacrificed to improve S/N on the other half.

The conventional Soller slit includes incident and outgoing beam masks that were wire-cut in 0.3 mm tungsten pate (calculated transmission $\sim 3 \times 10^{-9}$ at 25.7 keV). The incident mask was designed to be 68 mm from the sample position and had rectangular openings, 3.5 mm high by 0.32 mm wide on a ~0.830 mm pitch, while the outgoing beam mask was designed to be 177 mm from the sample, with the slit sizes enlarged proportionately. Foils of 0.1 mm thick Mo were used to separate each channel, and the location of these foils was determined by adjusting, under a microscope, the position of 6 (3 above, 3 below) stainless steel supports with wire-cut grooves to accept the foil (see fig. 4). Mo was chosen as it is both absorbing and relatively stiff. The entire assembly was then mounted on the XZθ assembly mounted on the spectrometer two-theta arm. This provides sufficient degrees of freedom to allow the Soller slit to be aligned with both the analyzers and the sample position. (Note that the vertical acceptance of the Soller slit, perpendicular to the scattering plane, was set generously so it did not have to be precisely aligned.) The collimator has been used to reduce backgrounds from windows and gas around liquid samples.

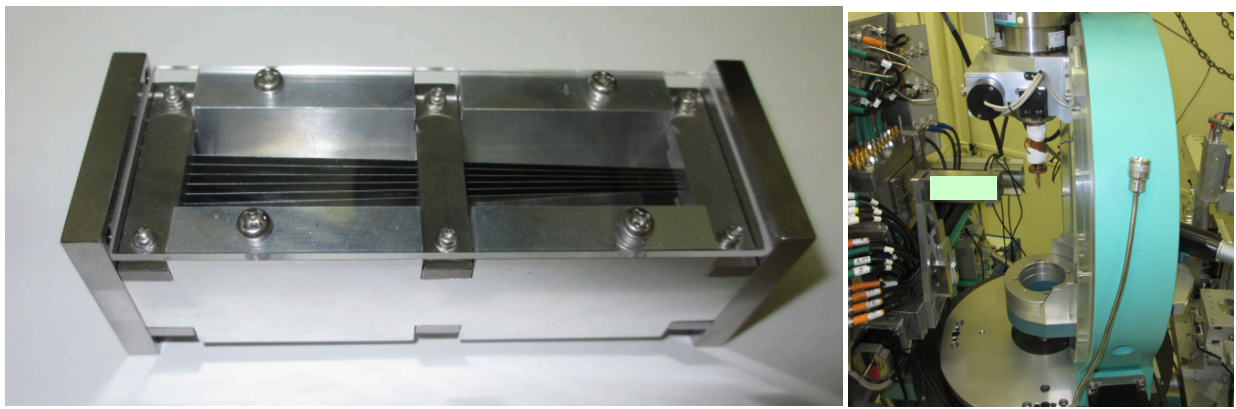

**FIGURE 4.** Conventional Soller slit separating analyzer channels in the scattering plane. The left panel is a close-up of the slit (approximately 115 mm long) while the right panel shows a photo of the sample area, including the Huber 512 cradle with a closed cycle cryostat mounted. The Soller Slit location is shown by the light rectangle just left of center.

Experiments on samples in diamond anvil cells (DACs) also could benefit from a Soller slit. However, here the constraints are more severe as one wants to reduce background either from the diamonds or, if used, a beryllium gasket, which are directly adjacent to the sample - typical dimensions along the x-ray beam path are two ~1.5 mm thick diamonds, or a ~6 mm diameter beryllium gasket. Meanwhile typical sample sizes are a few 10's of microns. Then the goal is then to separate the scattering from the central 10s of microns from that of the adjacent few mm. This requires that the entrance to the Soller slit be very close to the sample position and, consequently, that the Soller slit apertures have a very fine pitch - nominally 61 microns at 5 mm distance, as mentioned above. We were not successful in finding an appropriate fabrication technology that would allow the correct periodicity with the required tolerances for an acceptable price. However, we implemented a pair of screens (one 5 mm from the sample and one 10 mm further from the sample) on a common rigid support with apertures spaced by double the analyzer spacing, so 2x12.2 = 24.4 mrad, and *without* foils between the channels. By setting apertures of 0.048 mm H x 0.24 mm V on a 0.122 mm H pitch (laser cut into 0.1 mm tungsten, $4 \times 10^{-8}$ transmission at 17.79 keV) and a scaled aperture at 10 mm further from the sample (so apertures at 5 and 15 mm from the sample position), one can strongly cut the scattering from the diamonds and gasket, even without the foils - see fig. 5. However, this blocks the beam from every other analyzer, so comes at a cost, but, in some cases, cleaner data can be worth sacrificing half of the momentum transfers. The Soller screen assembly is mounted on the same XZθ assembly as used for the conventional Soller slit, and again, with a generous size in the vertical (transverse to the horizontal scattering plane) this is sufficient freedom to align the system with the analyzers and the sample position.

---
[*] Sometimes people also say "every second".

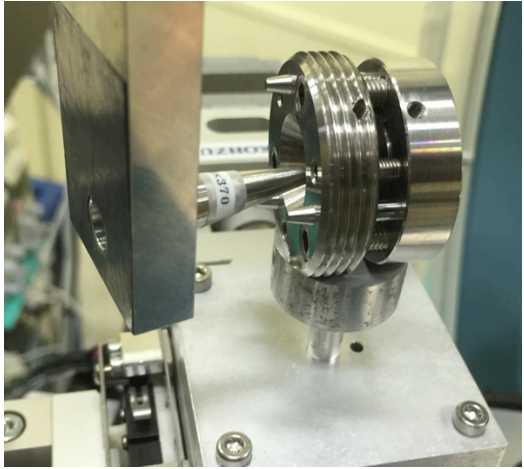 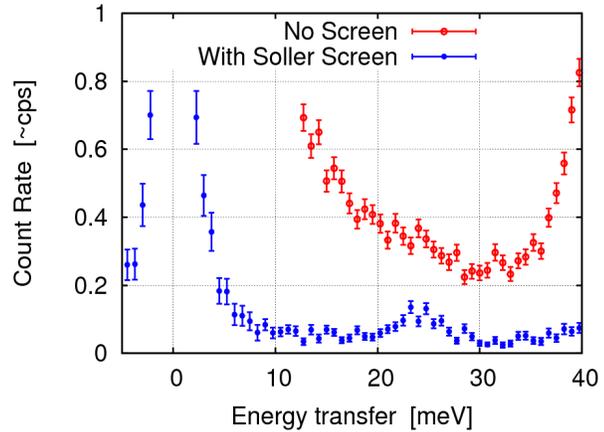

**FIGURE 5.** Left: Soller screen mounted near the DAC. Right: Spectra measured from a sample of Re metal at ~180 GPa pressure with and without the Soller screen. The reduction of the of the background from diamond and the improved contrast for the Re LA signal at 24 meV is clear.

## QUANDRANT BEAM POSITION MONITOR (BPM)

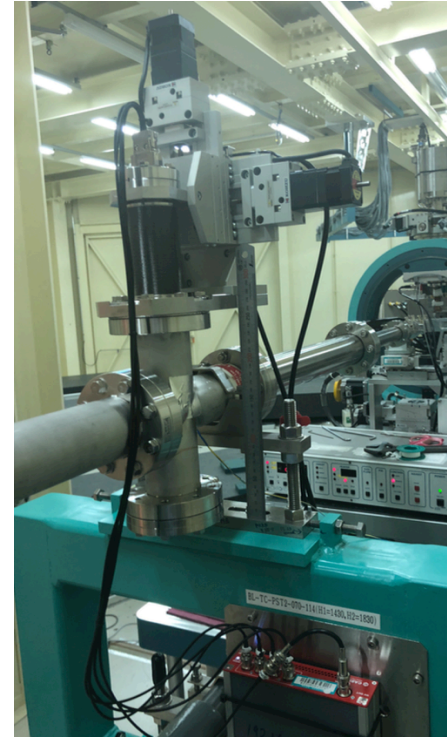

**FIGURE 6.** Diamond quadrant BPM mounted in a CF135 cross. X and Z stages are visible, as is the bellows for the motion feed-through. The current amplifier is at the bottom

A diamond beam-position monitor is installed 1.6 m upstream of the IXS sample location (see fig. 6). This uses a RIGI diamond quadrant sensor from Dectris held in rough (few Pa) vacuum, placed on an XZ stage with single bellows feed-through (conceptual design by Baron & Kohzu, fabrication by Kohzu), The electronics are read through a d-type connector coupled to a CAENels TetrAMM 4-channel current amplifier and bias unit read out over Ethernet. The CF135 cross is held on a steel support bolted to the experimental floor. Temperature monitoring of the support shows it is stable at the level of +-0.1 K (diurnal peak to valley) corresponding to motion of about +-3 microns for a coefficient of thermal expansion of ~15 ppm/K.

The x-ray beam position is calculated, in a linear approximation in the usual way from current differences divided by sums, with the scale factor determined from a small translation of the BPM, and a hard zero from the center of the quadrants. Up to issues of beam shape, this allows the beam position to be determined relative to the quadrant center over a range of about half of the FWHM of the beam. We use the BPM either in a partly focused beam (~0.3 mm) in our standard setup or in a large beam (1x3 mm$^2$) when the KB mirrors described above are used. The main purpose is to keep the beam position fixed during energy (temperature) scans of the backscattering monochromator (see [2]) which are often accompanied by concomitant drifts in the angle of the backscattering crystal: typically we set a threshold of 10 to 20 microns of beam motion *at the BPM* before it is corrected by adjusting the backscattering crystal stages. This fixes the position beam at the sample with a similar level of precision in the normal (~50 micron beam size) setup. When the KB is used, the placement of the BPM *before* the KB and the severe demagnification of the KB means the precision is better than 1 micron at the sample given the nearest optical element that changes the beam angle is ~18 m upstream of the BPM. The BPM then works nicely, even easily, with, perhaps, the most notable issues being proper grounding of the BPM and some software updates for the current amplifier. In general, the BPM provides ~micron resolution at the sensor location, even with the relatively weak, narrow bandwidth, x-ray beams used for IXS.

## CONCLUDING COMMENTS

We have discussed several "auxiliary" optical components that are useful for meV IXS with spherical analyzers. The multilayer KB performs well at 17.793 keV (2.8 meV resolution) and has become a standard part of the high-pressure beamline operation. A set optimized for 21.747 keV (1.3 meV resolution) should arrive within the year. The BPM setup is now in routine operation in all work. The other components have proven useful for specific experiments, with the masks and Soller slit used for most liquid measurements at low Q and the Soller screen used for some high pressure work, especially when beryllium gaskets are used.

## REFERENCES

[1] A. Q. R. Baron, SPring-8 Inf. Newsl. **15**, 14 (2010).
[2] A. Q. R. Baron, in *Synchrotron Light Sources Free. Lasers Accel. Physics, Instrum. Sci.*, edited by E. Jaeschke, S. Khan, J. R. Schneider, and J. B. Hastings (Springer International Publishing, Cham, 2016), p. 1643–1757. See also arXiv 1504.01098v4.
[3] D. Ishikawa, H. Uchiyama, S. Tsutsui, H. Fukui, and A. Q. R. Baron, in *Proc. SPIE - Int. Soc. Opt. Eng.* (2013).
[4] We use the term "venetian blind" slits to describe a set of slats (or flat plates) arranged in a geometry similar to venetian blind widow shades. The slits operate by rotating the slats about an axis centered in the middle of the slat to change the slit aperture. The axis locations are chosen so that when fully open the slats lie in the ~ 2mrad dead region between the analyzers and so do not occlude the beam, and when fully closed, the distance between the slat edges is ~2 mm. Arguably, this arrnagment could also be called persian blind slits, or slat-type slits, but here we use venetian as being a common/familiar choice.
[5] W. Soller, Phys. Rev. **24**, 158 (1924).